\documentclass[fleqn,twoside]{article}
\usepackage{gc}
\renewcommand{\Authors}[4]{\noi
        {\large\bf #1\dag\ddag\ #2\ddag}\medskip\begin{description}
        \item[\dag]{\it #3} \item[\ddag]{\it #4}\end{description}}
\newcommand{\la}[1]{\label{#1}}
\newcommand{\ur}[1]{(\ref{#1})}

\newcommand{\urss}[3]{(\ref{#1}), (\ref{#2}), (\ref{#3})}

\renewcommand{\eq}[1]{Eq.\,(\ref{#1})}
\renewcommand{\eqs}[2]{Eqs.\,(\ref{#1}), (\ref{#2})}

\newcommand{\ee}[1]{e_{\{#1\}}}

\def\Tr{\mbox{Tr}\,}
\def\bea{\begin{eqnarray}}
\def\eea{\end{eqnarray}}

\heads{D.Diakonov and V.Petrov}
	{Yang--Mills theory as a quantum gravoty with `\ae ther'}

\begin{document}
\twocolumn[
\Arthead{8}{2002}{1/2 (29/30)}{1}{11}

\Title{YANG--MILLS THEORY AS A QUANTUM GRAVITY WITH `\AE THER'}

\Authors{Dmitri Diakonov\foom 1} {and Victor Petrov\foom 2}
	{NORDITA, Blegdamsvej 17, DK-2100 Copenhagen, Denmark}
	{St. Petersburg Nuclear Physics Institute, Gatchina 188300, Russia}

\Rec{}

\Abstract
{Quantum Yang--Mills theory can be rewritten in terms of gauge-invariant
variables: it has the form of the so-called BF gravity, with an additional 
`\ae ther' term. The BF gravity based on the gauge group $SU(N)$ is actually a
theory
 of high spin fields (up to $J=N$) with high local symmetry mixing up
fields
 with different spins --- as in supergravity but without fermions. As
$N\to\infty$, one gets a theory with an infinite tower of spins related by
local symmetry, similar to what one has in string theory. We thus outline
a way of deriving a string theory from the local Yang--Mills theory in the
large $N$ limit.  }

\vspace{40mm}

] 
\email 1 {diakonov@nordita.dk}
\email 2 {victorp@thd.pnpi.spb.ru}

\section{Introduction}

It is widely believed that the Yang--Mills theory in the large $N$ limit
is equivalent to some version of string theory \cite{AMP1,MM}. Recent
attempts to justify this equivalence from the string side have been
described by A. Polyakov \cite{AMP2}. In this paper, we propose how to
derive string theory starting from the side of local field theory.

The apparent difficulty of this programme is that the Yang--Mills theory is
formulated in terms of local gauge-noninvariant connection $A_\mu$
whereas string theory is formulated in terms of gauge-invariant although
nonlocal variables. Therefore, a natural first step to link the two
approaches is to rewrite the Yang--Mills theory in terms of gauge-invariant
variables. It can be done via the first order formalism \cite{DT,Halpern}
by introducing an additional Gaussian integration over dual field
strength and then integrating out the Yang--Mills connection $A_\mu$.
That is always possible since the integral over $A_\mu$ is Gaussian. The
remaining action depending on the dual field strength can be organized in
such a way that only gauge-invariant variables appear. One of the goals of
the paper is to demonstrate that the Yang--Mills theory can be reformulated
in terms of gauge-invariant variables. We do it for the $SU(2)$ gauge group
in $d=4$ and for arbitrary $SU(N)$ in $d=3$.

An intriguing feature of the first-order formalism is that it reveals a
hidden invariance of the Yang--Mills theory (in flat space!) with respect to
general coordinate transformations, or diffeomorphisms. This has been
first noticed by Lunev \cite{Lun1} in the simplest case of the $SU(2)$
Yang--Mills theory in $d=3$ and later on extended by Ganor and
Sonnenschein to $d=4$ \cite{GS}. The gravitational parallel refers to
the `colour dual space' which is curved. We explain the origin of the
diffeomorphism invariance in the next section. To be more precise, only
one term of the action in the first-order formalism (the `mixed' term)
possesses this symmetry, the other term does not but it is a trivial one
(we call it the `\ae ther' term). If the `\ae ther' term is neglected
the theory is known under the name of BF gravity; it is a topological field
theory, see the reviews \cite{BFobzor,BFgrav}.

The BF action has enormous local symmetry: it is invariant both under
`normal' and `dual' gauge transformations. Both symmetries are apparent when
the action is presented in terms of the Yang--Mills connection $A_\mu$.
However, we need to integrate out the connection. The symmetry under
`normal' gauge transformations is automatically taken into account when one
writes the result in terms of gauge-invariant variables, but what about the
symmetry under dual gauge transformations? To our knowledge, this question
has not been scrupulously addressed before. The main finding of this paper
is that dual gauge invariance translates into an exciting new symmetry,
namely a local symmetry under mixing fields with different spins.

This symmetry can be parallelled to supergravity which is invariant under
local rotations of fields carrying integer and half-integer spins, and to
string theory where an infinite tower of spins is related by an
infinite-dimensional algebra. In contrast to the former theory, we have
only boson fields, and the number of higher spins can be arbitrary. In
contrast to the latter theory, we can have a finite number of higher spins.
Only in the limit $N\to\infty$ of the $SU(N)$ gauge group the symmetry
relates an infinite tower of spins. However, it is only in the large $N$
limit that the Yang--Mills theory is expected to be equivalent to string
theory.

The paper is organized as follows. In the next section we explain why the
`mixed' term of the first order formalism is diffeomorphism-invariant. In
\sect 3 we write down the $SU(2)$ Yang--Mills theory in $d=4$ in terms of
gauge-invariant variables.  We indicate how, despite it, one recovers gluons
at short distances. From the gravity point of view, we rewrite the BF
action in terms of basis-independent variables. We reveal the 12-function
invariance of the BF gravity and show that it mixes fields with different
spin content.

Starting from \sect 4, we concentrate on $d=3$ theories where we are able
to go further than in $d=4$. In \sect 5 we briefly recall the solution for
the $SU(2)$ case. In \sect 6 we proceed to higher $SU(N)$ groups and
introduce gauge-invariant variables. These turn out to be fields carrying
spin from zero to $N$; each spin appears twice, except the `edge' spins
0,\ 1,\ $N-1$ and $N$, which appear only once. We point out the
transformation of those spins through one another, which leaves the BF
action invariant. Finally, we speculate that the noninvariant `\ae ther'
term lifts the degeneracy of spins and gives rise to a finite string slope
$\alpha'$.

\section{Hidden diffeomorphism invariance of Yang--Mills theory in flat
	 space}

The key observation is the following. Let us rewrite identically the
partition function of the Yang--Mills theory in flat Euclidean $d=4$
space with the help of an additional Gaussian integration over dual
field strength variables \cite{DT,Halpern}:
\bea
  {\cal Z} \eql \int DA_\mu\,\exp\int d^4x\;\left(- \frac{1}{2g^2}\Tr\,
		F_{\mu\nu}F_{\mu\nu}\right)
\nnn \quad
      \int DA_\mu\,DG_{\mu\nu}\,\exp\int
	d^4x\; \biggl( -\frac{g^2}{2} \Tr\, G_{\mu\nu}G_{\mu\nu}
\nnn \inch 	\la{Z1}
	+ \frac{i}{2}\,
    \epsilon^{\alpha\beta\mu\nu}\,\Tr\,G_{\alpha\beta}F_{\mu\nu}\biggr).
\eea
where $F_{\mu\nu}=\d_\mu A_\nu-\d_\nu A_\mu-i[A_\mu A_\nu]$ is
the standard Yang--Mills field strength and $\epsilon^{\alpha\beta\mu\nu}$
is the antisymmetric tensor. To avoid possible confusion we write down
explicitly all indices. To be specific, the gauge group is $SU(N)$ with
$N^2-1$ generators $t^a,\quad \Tr\,t^at^b=\delta^{ab}/2$.  \eq{Z1} is
called the first-order formalism.

Both terms in \eq {Z1} are invariant under the $(N^2-1)$-function gauge
transformation
\bear
	\delta A_\mu \eql[{\cal D}_\mu\, \alpha],
\nn
	\delta G_{\mu\nu} \eql [G_{\mu\nu}\,\alpha],
\la{GT}\ear
where ${\cal D}_\mu=\d_\mu -iA^a_\mu t^a$ is the Yang--Mills
covariant derivative, $[{\cal D}_\mu {\cal D}_\nu]=-iF_{\mu\nu}$.

Due to the Bianchi identity, $\epsilon^{\mu\nu\rho\sigma}\,[{\cal D}_\nu
F_{\rho\sigma}]=0$, the second (mixed) term in \eq{Z1} is, in addition,
invariant under the $4\cdot(N^2-1)$-function `dual' gauge transformation,
\bear
	\delta A_\mu \eql  0,
\nn
	\delta G_{\mu\nu} \eql
	[{\cal D}_\mu\beta_\nu]-[{\cal D}_\nu\beta_\mu].	 \la{DGT}
\ear
Taking a particular combination of the functions in \eqs {GT} {DGT},
\beq
    \alpha=v^\mu A_\mu,\qquad \beta_\mu=v^\lambda G_{\lambda\mu}, \la{CT}
\eeq
leads to the transformation
\beq
	\delta G_{\mu\nu}= -G_{\lambda\nu}\,\d_\mu v^\lambda
		-G_{\mu\lambda}\,\d_\nu v^\lambda -\d_\lambda
				G_{\mu\nu}\,v^\lambda, \la{CTT}
\eeq
being the known transformation of a (covariant) tensor under general
coordinate transformation. Therefore, the `mixed' term is
diffeomorphism-invariant, and is known as BF gravity\footnote
	{With our notations it would be more appropriate to call it `GF
	gravity' but we follow the tradition.}.
It defines a topological field theory of Schwarz type \cite{BFobzor}.
Moreover, it is invariant not under four but under as much as $4\cdot
(N^2-1)$ local transformations; four diffeomorphisms are but their small
subset. We shall see later on that the additional local transformations mix
up fields with different spins.

The first term in \eq{Z1} is not invariant under the dual gauge
transformation \ur{DGT}, therefore it is not invariant under
diffeomorphisms. For that reason, we call it the `\ae ther' term:
it distinguishes the Yang--Mills theory from a non-propagating
topological BF gravity represented by the second (mixed) term in the
action \ur{Z1}.

\section{$SU(2)$, $d=4$ BF gravity in a basis-independent formulation}

In the first-order formalism, the integral \ur{Z1} over the Yang--Mills
connection $A_\mu$ is Gaussian, and one can integrate it out. This was
done many years ago \cite{DT,Halpern}, but in contrast to that work
we wish to write down the result of the integration in an explicitly
gauge-invariant way. This was performed some time ago by Ganor and
Sonnenschein \cite{GS}; it has been shown that the resulting theory contains
the Einstein--Hilbert action for the metric tensor and an additional
5-component self-dual field interacting with the metric. The final action of
Ref.\,\cite{GS} is very lengthy, and its symmetry under a 12-function
transformation has not been discussed. However, an important finding of
\Ref{GS} is the way one constructs the metric tensor out of the dual
field strength $G_{\mu\nu}$.

In this section, we write down the result of the $A_\mu$ integration
in a compact form which makes clear the 12-function symmetry of the BF
action. From the point of view of the Yang--Mills theory, it solves the
problem of reformulating it in terms of local gauge-invariant variables.
From the point of view of BF gravity, we rewrite it in a basis-invariant
formalism.

The Gaussian integration over $A_\mu$ in \eq{Z1} is equivalent to the
saddle-point approximation. The saddle point (which we denote by $\bar
A_\mu$) is found from varying the `mixed' term in $A_\mu$:
\beq
	\epsilon^{\lambda\mu\alpha\beta}\,
		[{\cal D}_\mu(\bar A)\,G_{\alpha\beta}]=0.
\la{UD4}\eeq
We need to solve this equation with respect to the saddle-point YM
connection $\bar A_\mu$ and to substitute it back into the BF action
\beq
	S_2=\frac{i}{2}\int d^4x\,
	\epsilon^{\alpha\beta\mu\nu}\,\Tr\,G_{\alpha\beta}F_{\mu\nu}(\bar A).
\la{S2}\eeq
The goal is to write down the result for $S_2$ through gauge-invariant
combinations made of the dual field strength $G_{\alpha\beta}$ and to reveal
its 12-function symmetry.

\subsection{Gauge-invariant variables}

First of all, we need a convenient parametrization of $G^a_{\alpha\beta}$
which have $6\cdot 3$ degrees of freedom (dof's), out of which $18-3=15$ are
gauge-invariant. Our main variable will be an antisymmetric tensor
$T^i_{\alpha\beta}=-T^i_{\beta\alpha}$ (the Greek indices run from 1 to 4
whereas the Latin ones run from 1 to 3). Given $T$, one constructs the
quantity
\bearr
	(\sqrt{g})^3 \eqdef \frac{1}{48}\left(
  \epsilon_{ijk}\,T^i_{\alpha\beta}T^j_{\gamma\delta}T^k_{\epsilon\eta}
  \right)\;(\epsilon_{lmn}\,T^l_{\kappa\lambda}T^m_{\mu\nu}T^n_{\rho\sigma})
\nnn \cm\cm
	\times\epsilon^{\alpha\beta\kappa\lambda}
		\epsilon^{\gamma\delta\mu\nu}\,
			\epsilon^{\epsilon\eta\rho\sigma}. \la{sqrtg}
\ear
With its aid we construct the contravariant antisymmetric tensor
\beq
	T^{i\,\mu\nu}\eqdef\frac{1}{2\sqrt{g}}\,\epsilon^{\mu\nu\alpha\beta}\,
		T^i_{\alpha\beta} \la{Tcontra}
\eeq
and require the orthonormalization condition,
\beq
    T^i_{\alpha\beta}T^{j\,\alpha\beta}=\delta^{ij}.             \la{orthon}
\eeq
This condition is `dimensionless' in $T$, therefore it imposes 5 rather than
6 constraints on the covariant tensor $T^i_{\alpha\beta}$, which thus
carries $18-5=13$ dof's. The general solution to \eq{orthon} is given by
\beq
	T^i_{\alpha\beta}=\eta^i_{AB}\,e^A_\alpha\,e^B_\beta   \la{realiz}
\eeq
where $e^A_\alpha$ can be called a tetrad; $\eta^i_{AB}$ is
the 't Hooft symbol whose algebra is given in \cite{tH}. All algebraic
statements of this section can be verified by exploiting the
$\eta$-symbol algebra. There are 16 dof's in the tetrad, however three
rotations under one of the $SO(3)$ subgroups of the $SO(4)$ Euclidean
group do not enter into the combination \ur{realiz}, therefore the
r.h.s. of \eq{realiz} carries, as it should, 13 dof's.

We next introduce the metric tensor,
\beq
	g_{\mu\nu}\eqdef\frac{1}{6}\,\epsilon^{ijk}\,T^i_{\mu\alpha}
		T^{j\,\alpha\beta}T^k_{\beta\nu}=e^A_\mu\,e^A_\nu.
\la{metr4}\eeq
It explains the previous notation: \eq{sqrtg} is consistent with the
determinant of this metric tensor.

Finally, we parametrize the dual field strength as
\beq
	G^a_{\alpha\beta}=d^a_i\,T^i_{\alpha\beta}=d^a_i\,\eta^i_{AB}\,
		e^A_\alpha\,e^B_\beta,
\la{Gmunu}\eeq
where the new variable $d^a_i$ (we shall call it a triad) is subject
to the normalization constraint $\det\,d^a_i= 1$ and therefore contains
8 dof's. In fact, the combination \ur{Gmunu} is invariant under
simultaneous $SO(3)$ rotations of $T^i$ and $d_i$, therefore the r.h.s.
of \eq{Gmunu} contains $13+8-3=18$ dof's, as does the l.h.s. Thus,
\eq{Gmunu} is a complete parametrization of $G^a_{\mu\nu}$.

It is now clear how to organize the 15 gauge-invariant variables made of
$G^a_{\mu\nu}$. These are the 5 dof's contained in the symmetric $3\times
3$ tensor
\beq
	h_{ij}\eqdef d^a_i\,d^a_j,\qquad \det\, h=1,
\la{defh}\eeq
and 13 dof's of $T^i_{\alpha\beta}$. However, $h_{ij}$ and
$T^i_{\alpha\beta}$ will always enter contracted in $i,j$ (as follows from
\eq{Gmunu}), so that the dof's associated with the simultaneous $SO(3)$
rotation will drop out. In other words, one can choose $h_{ij}$ to be
diagonal and containing only 2 dof's.

\subsection{Christoffel symbols, covariant derivative, Riemann tensor}

We are now prepared to solve the saddle-point equation (\ref{UD4}) and to
express the BF action \ur{S2} in a nice geometric way. We substitute
\ur{Gmunu} into \eq{UD4} and rewrite it as
\bearr \nhq
   0=\frac{1}{2}\epsilon^{\lambda\mu\alpha\beta}\,D^{ab}_\mu(\bar A)\,
    (d^b_i\,T^i_{\alpha\beta}) = \frac{1}{\sqrt{g}}
     D^{ab}_\mu (d^b_i\,\sqrt{g}\, T^{i\,\lambda\mu})
\nnn \cm
	=\frac{1}{\sqrt{g}}\d_\mu
		(\sqrt{g}\,T^{i\,\lambda\mu})d^a_i
			+T^{i\,\lambda\mu}\,D^{ab}_\mu d^b_i.
\la{UD41}\ear
The action of the YM covariant derivative on the triad can be
decomposed in the triad again:
\beq
		D^{ab}_\mu\,d^b_i=\gamma^j_{\mu i}\,d^a_j
\la{mChrdef}\eeq
which serves as a the definition of the `minor' Christoffel symbol
$\gamma^j_{\mu i}$ (not to be confused with the ordinary Christoffel
symbol in $d=4$). With its help we define the `minor' covariant derivative,
\beq
	(\nabla_\mu)^j_i  \eqdef \d_\mu\,\delta^j_i + \gamma^j_{\mu i}
\la{mcovder}\eeq
and the `minor' Riemann tensor,
\bearr                           \nq
   R^j_{\;i\,\mu\nu}\!=\!\left[\nabla_\mu\,\nabla_\nu\right]^j_i
       \!=\!\d_\mu\gamma^j_{\nu i}-\d_\nu\gamma^j_{\mu i}
	+\gamma^j_{\mu k}\gamma^k_{\nu i}-\gamma^j_{\nu k}\gamma^k_{\mu i}\!.
\la{mRiemann}                            \nnn
\ear
The saddle point equation (\ref{UD41}) can be compactly written as
\bear
	(\nabla_\mu)^j_i (\sqrt{g}\,T^{i\,\lambda\mu})\eql 0,
\quad{\rm or} \nn
	T_{\kappa\lambda;\;\mu}+T_{\lambda\mu;\;\kappa}+
			T_{\mu\kappa;\;\lambda}\eql 0,
\la{covderTz}
\eea
meaning that the antisymmetric tensor $T^{\alpha\beta}$ is `covariantly
constant'.  Another consequence of \eqs{UD41}{mChrdef} is that the
symmetric tensor $h_{ij}$ is covariantly constant, too:
\beq
	h_{ik;\;\mu}\eqdef \d_\mu\,h_{ik}-\gamma^j_{\mu i}\,h_{kj}
		-\gamma^j_{\mu k} h_{ij}=0.
\la{covderhz}\eeq
The `minor' Christoffel symbol can be found explicitly; it consists of
symmetric and antisymmetric parts:
\bear
\gamma^j_{\mu i}\eql  \half h^{jn}(\d_\mu
			h_{ni}+\epsilon_{nik}\,S^k_\mu),
\nn
	S^k_\mu \eql T^k_{\nu\beta}T^l_{\mu\alpha}g^{\alpha\beta}
			\Bigl [h_{lm}\d_\lambda T^{m\,\lambda\nu}
\nnn \cm
	+\frac{1}{2g} T^{m\,\lambda\nu}\d_\lambda(g\,h_{lm})\Bigr],
\la{mChr1}
\ear
where we have used contravariant upper indices to denote the inverse
matrices $h^{jn},\ g^{\alpha\beta}$.

Given the Christoffel symbol, one may return to \eq{UD41} and find
the saddle-point YM field $\bar A_\mu$: it coincides with the old
result of Refs.\,\cite{DT,Halpern}. However, we do not need an explicit
form of $\bar A_\mu$ to find the action \ur{S2} at the saddle point.

\subsection{Action in terms of gauge-invariant variables}

In order to find the Yang--Mills field strength $F_{\mu\nu}$ at the
saddle point we consider the double commutator of YM covariant derivatives,
\beq
   [{\cal D}_\mu[{\cal D}_\nu d_i]]=[{\cal D}_\mu,\, \gamma^j_{\nu i}d_j]
   =d_j(\d_\mu\gamma^j_{\nu i}+\gamma^j_{\mu k}\gamma^k_{\nu i}),
\la{doublecomm}\eeq
and subtract the same commutator with $(\mu\nu)$ interchanged:
\bearr
   [{\cal D}_\mu[{\cal D}_\nu d_i]]-[{\cal D}_\nu[{\cal D}_\mu d_i]]
		=-[d_i[{\cal D}_\mu{\cal D}_\nu]]
\nnn \cm
		=i[d_i\,F_{\mu\nu}(\bar A)]=d_j\,R^j_{\;i\,\mu\nu}.
\la{Fmunu}\ear
Hence we see that the YM curvature at the saddle point is expressed
via the `minor' Riemann tensor, \eq{mRiemann}. Explicitly,
\beq
	F^a_{\mu\nu}(\bar A)=\half
		\epsilon^{abc}\,d^{bi}\,d^c_j\,R^j_{\;i\,\mu\nu}
\la{Fmunu1}\eeq
where the inverse triad, $d^{bi}d^b_l=\delta^i_l$, has been used;
$d^{bi}=h^{im}d^b_m$. We put \eq{Fmunu1} into the action \ur{S2} and
get finally
\beq
       S_2=\frac{i}{4}\int d^4x\,\sqrt{g}\,R^j_{\;i\,\mu\nu}\,T^{l\,\mu\nu}\,
		\epsilon_{jlk}\,h^{ki}. \la{act42}
\eeq
This is the $SU(2)$, $d=4$ BF gravity action in the gauge-invariant or
basis-independent formulation, since it is expressed in terms
of the gauge-invariant variables $T$ and $h$.  We notice that it is
covariant with respect to both Greek and Latin indices.

To get the full YM action in a gauge-invariant form one has to add the first
(`\ae ther') term of \eq{Z1},
\beq
	S_1=-\frac{g^2}{4}\int d^4x\, T^i_{\mu\nu}\,h_{ij}\,T^j_{\mu\nu},
\la{act41}\eeq
which is gauge- but not diffeomorphism-invariant.

In the particular case when $h_{ij}=\delta_{ij}$, the action \ur{act42} can
be rewritten in terms of the 4-dimensional metric tensor $g_{\mu\nu}$ being
a particular combination of $T^i_{\mu\nu}$, see \eq{metr4}. In this case the
BF action \ur{act42} becomes the usual Einstein--Hilbert action,
\beq
	S_2\Big|_{h_{ij}=\delta_{ij}}=\frac{i}{2}\int d^4x\,\sqrt{g} R,
\la{EH}\eeq
where $R$ is the standard scalar curvature made of $g_{\mu\nu}$.

Another particular case is an arbitrary (but constant) field $h_{ij}$
and a conformally flat metric, $g_{\mu\nu}=\Phi\,\delta_{\mu\nu}$. In
this case (being of relevance to instantons) the BF action \ur{act42} is
\bearr
	S_2\Big|_{\rm conf.flat} = \frac{i}{2}\int d^4x\,
	\left(-\d^2\Phi+\frac{1}{2\Phi}\d_\lambda\Phi\d_\lambda\Phi\right)
\nnn \inch
		\times (h_{ii}h_{jj}-2h_{ij}h_{ij}).	\la{confflat}
\eea

Our choice of the gauge-invariant variables $T,\ h$ is not imperative.
For example, one can use the 15 variables
\beq
	W_{\alpha\beta\gamma\delta}\eqdef
	G^a_{\alpha\beta}G^a_{\gamma\delta}=
	T^i_{\alpha\beta}\,h_{ij}\, T^j_{\gamma\delta}
\la{W}\eeq
or some other set of 15 variables, depending on what properties of the
theory one wishes to fix upon.

\subsection{Gauge-invariant perturbation theory}

A seeming paradox is that the Yang--Mills theory has gluon degrees of
freedom at short distances, whereas in a gauge-invariant formulation
there is no place for explicitly colour degrees of freedom. We shall show now
that \eqs{act42}{act41} possess two transversely polarized gluons
(times 3 colours). This is the correct gauge-invariant content of the
perturbation theory at zero order.

Since $S_1$ is proportional to the coupling constant and $S_2$ is not,
the zero order corresponds to $S_2=0$, i.e., to the `minor' Riemann tensor
$R^i_{\;j\,\mu\nu}=0$, that is to the flat dual space. It implies that the
`minor' Christoffel symbol $\gamma_{\mu j}^i$ is a ``pure gauge'',
\beq
	\gamma_{\mu j}^i=\left(O^{-1}\right)^i_k\d_\mu O^k_j,
			\qquad \det\,O\neq 0.
\la{pg2}\eeq
Indeed, in this case the Riemann tensor is zero:
\bear
	(\nabla_\mu)^i_jc^j\eql \d_\mu c^i+ \gamma_{\mu j}^ic^j
		=(O^{-1})^i_k\d_\mu (O^k_jc^j);
\nnv
	R^i_{\;j\,\mu\nu}c^j
	\eql \left[(\nabla_\mu)^i_k  (\nabla_\nu)^k_j
			-(\mu\leftrightarrow\nu)\right] c^j
\nn \nq   \la{zeroR}
	\eql (O^{-1})^i_l\d_\mu
		\d_\nu (O^l_jc^j) - (\mu\leftrightarrow\nu)=0
\eea
for any vector $c^j$, therefore $R^i_{\;j\,\mu\nu}=0$.

We substitute \eq{pg2} into \eq{covderhz} and get
\beq
	0=h_{ik;\;\mu}=\d_\mu\left[\left(O^{-1}\right)^p_lh_{pq}
	\left(O^{-1}\right)^q_m\right]\;O^m_iO^l_k
\la{11}\eeq
meaning that $h_{pq}=O^i_pO^j_q\, D_{ij}$ where $D_{ij}$ is a constant
matrix. Next, we substitute \eq{pg2} into \eq{covderTz} and obtain
\beq
	\d_\kappa\left(O^k_jT^j_{\lambda\mu}\right)
	+\d_\lambda\left(O^k_jT^j_{\mu\kappa}\right)
	+\d_\mu\left(O^k_jT^j_{\kappa\lambda}\right) = 0,
\la{21}\eeq
whose general solution is $O^k_jT^j_{\kappa\lambda} =\d_\kappa B^k_\lambda
-\d_\lambda B^k_\kappa$. The first term in the action ($S_1$) is then
\bear
    G^a_{\mu\nu}G^a_{\mu\nu} \eql  h_{pq}T^p_{\mu\nu}T^q_{\mu\nu}
			=D_{ij}(O^i_pT^p_{\mu\nu})(O^j_qT^q_{\mu\nu})
\nn            \la{trgluons}
\eql D_{ij} \left(\d_\mu B^i_\nu-\d_\nu B^i_\mu\right)\!
	    \left(\d_\mu B^j_\nu-\d_\nu B^j_\mu\right)\!.
\eea
$D_{ij}$ is a constant matrix and can be set to be $\delta_{ij}$ by a linear
transformation of the three vector fields $B_\mu^i$; therefore, we obtain
the Lagrangian of three massless gauge fields. It is an expected result.

In Ref.\,\cite{MZ}, the correct renormalization of the gauge coupling
constant has been demonstrated in the first-order formalism. It would be
most instructive to follow how ``11/3'' of the Yang--Mills
$\beta$ function arises in the gauge-invariant formulation.

\subsection{More general relativity}

As discussed in \sect 2, the BF action $S_2$ is invariant under 12-function
dual gauge transformations, 4 of which are the general coordinate
transformations or diffeomorphisms. In this subsection we describe how this
12-function symmetry is translated after one integrates out the YM
connection and arrives at the gauge-invariant action \ur{act42}.

The dual gauge transformation \ur{DGT} can be written as
\beq
	\delta G^a_{\mu\nu}=D^{ab}_\mu\beta^b_\nu-D^{ab}_\nu\beta^b_\mu.
\la{beta}\eeq
We decompose 12 functions $\beta^b_\mu$ in the triad basis,
\beq
		\beta^b_\mu=z^i_\mu d^b_i
\la{z}\eeq
where $z^i_\mu$ is another set of 12 arbitrary infinitesimal functions.
Putting \eq{z} into \eq{beta} and using \eq{mChrdef}, we obtain the variation
\bear               \la{p1}
	\delta G^a_{\mu\nu}\eql d^a_i\,p^i_{\mu\nu},
\\      		\la{p2}
	p^i_{\mu\nu}\eql (\nabla_\mu)^i_j z^j_\nu-(\nabla_\nu)^i_j z^j_\mu,
\eea
where $\nabla_\mu$ is the covariant derivative \ur{mcovder}. Importantly, we
have excluded the YM connection $\bar A_\mu$ from the variation by using the
saddle-point equation \ur{mChrdef}. The variation \ur{p1} is written in
terms of the variables entering into $S_2$ {\em after} the Gaussian
integration over $A_\mu$ is performed. Therefore, $S_2$ in the resulting
form of \eq{act42} is, by construction, invariant under the 12-function
variation \ur{p1}.

We need to derive the transformation laws for the
gauge-invariant quantities $g_{\mu\nu}, T^i_{\mu\nu}, h_{ij}$,
that follow from \eq{p1}. First of all, we find the transformation law
for the metric tensor \ur{metr4} which can be written as \cite{GS}
\beq
	g_{\mu\nu}=\frac{1}{6}\epsilon^{abc}\,\frac{\epsilon^{\alpha\beta\rho
	\sigma}} {2\sqrt{g}}\,G^a_{\mu\alpha}G^b_{\rho\sigma}G^c_{\beta\nu}.
\la{metr41}
\eeq
The variation of the metric tensor under the transformation \ur{p1} is
\bea                                         		\la{varg}
	\delta g_{\mu\nu}\eql \frac{1}{2}\left(\!
	g_{\mu\beta}p^i_{\alpha\nu}+g_{\beta\nu}p^i_{\alpha\mu}-\frac{1}{3}
	g_{\mu\nu}p^i_{\alpha\beta}\!\right)T^{i\,\alpha\beta}\!,
\\  \la{vardetg}
    \frac{\delta g}{g}\eql \frac{1}{3}\,p^i_{\alpha\beta}T^{i\,\alpha\beta}.
\eea
The variation of the covariant tensor $T^i_{\mu\nu}$ is
\bear                                                         \la{varTcov}
	\delta T^k_{\mu\nu}
	\eql p^k_{\mu\nu}-Q^k_i\,T^i_{\mu\nu},
\\    \la{OqQ}
	Q^k_i \eql \fract{1}{4}\left(\delta^l_i\delta^k_m-\fract{1}{3}
       \delta^k_i\delta^l_m\right)p^l_{\alpha\beta}T^{m\,\alpha\beta},
	\quad Q^i_i = 0.
\eea
The variation of the contravariant tensor $T^{i\,\mu\nu}$ can be found from
\beq
	\delta T^{k\,\mu\nu}=
	\frac{\epsilon^{\mu\nu\kappa\lambda}}{2\sqrt{g}}\,
	\left(\delta T^k_{\kappa\lambda}-\frac{1}{2}\frac{\delta g}{g}\,
	T^k_{\kappa\lambda}\right),
\la{varTcontra}\eeq
which supports the self-duality property, \eq{Tcontra}.

Finally, we find the transformation of $h_{ij}$ to be
\beq
		\delta h_{ij}=Q^k_i\,h_{kj}+h_{ik}\,Q^k_j,
\la{varh}\eeq
which supports $\det\,h$=1 under the variation. It should be noted
that the variations \urss{varTcov}{varTcontra}{varh} are written up to
possible $SO(3)$ rotations in the Latin indices.

A special 4-function subclass of transformations are diffeomorphisms.
This particular set of transformations is obtained by choosing in \eq{z}
\beq
		z^i_\mu=T^i_{\mu\lambda}\,v^\lambda,
\la{zdiff}\eeq
where $v^\lambda$ is the infinitesimal displacement vector,
$x^\lambda\to x^\lambda-v^\lambda(x)$. It corresponds to taking
\beq
		p^i_{\mu\nu}=\d_\mu v^\lambda\,T^i_{\nu\lambda}
		-\d_\nu v^\lambda\,T^i_{\mu\lambda}
	-v^\lambda\left(\nabla_\lambda\right)^i_jT^j_{\mu\nu}.
\la{pdiff}\eeq

It is a matter of simple algebra to verify that on this subclass the
variation of the metric tensor \ur{varg} becomes
\beq
	\delta g_{\mu\nu}\Big |_{{\rm diff}}
	=-g_{\mu\lambda}\,\d_\nu v^\lambda-g_{\lambda\nu}\,\d_\mu
			v^\lambda -\d_\lambda g_{\mu\nu}\, v^\lambda,
\la{vargdiff}\eeq
which is the usual transformation under diffeomorphisms. Similarly, one
finds that under general coordinate transformations $T^i_{\mu\nu}$
transforms as a covariant tensor, while $h_{ij}$ is a world scalar,
\beq
	\delta h_{ij}\Big|_{\rm diff}=-\d_\lambda h_{ij}\,v^\lambda.
\la{varhdiff}\eeq

In general, however, the BF action \ur{act42} is invariant not only under
4-function diffeomorphisms but under full 12-function transformations
described above. The additional 8-function transformations mix, in a
nonlinear way, the fields $h_{ij}$ and $T^i_{\mu\nu}$, i.e., world scalars
with covariant tensors. In other words, the BF action has a large local
symmetry which mixes fields with different spin content.

This symmetry is, of course, a consequence of the invariance of the
BF action under the dual gauge transformations \ur{DGT}: it reveals itself
when one integrates out the Yang--Mills connection $A_\mu$. The number
of free functions determining the dual gauge transformation is
$4\cdot(N^2-1)$ for the $SU(N)$ gauge group; for $SU(2)$ it is 12. For
higher groups there will be more degrees of freedom in the symmetry.
Simultaneously, for higher groups the BF action will involve higher spin
fields, and the invariance under dual gauge transformation will be
translated, after excluding the YM connection, into the invariance
under mixing higher spins.

So far we have not developed the BF theory for higher groups in $d=4$ but
only in $d=3$ where the formalism is simpler \footnote{There have been
suggestions how to treat the $SU(N)$ gauge group in $d=4$ \cite{Lun2}, but
we follow another route in this paper.}. Therefore, in the rest of the paper
we concentrate on the gauge-invariant formulation of the Yang--Mills theory
in $d=3$. We shall show that when one integrates out the Yang--Mills
connection from the BF action, one obtains a theory of fields carrying spin
up to $J=N$ (for the $SU(N)$ gauge group), and that the invariance under
dual gauge transformations is translated into a local symmetry which mixes
all those spins.

\section{First-order formalism in $d=3$}

In Euclidean $d=3$ dimensions the Yang--Mills partition
function in the first-order formalism reads:
\bearr
	{\cal Z} = \int DA^a_i\;
	     \exp\left(-\frac{1}{4g_3^2}\int d^3x\;F^a_{ij}F^a_{ij}\right)
\nnn
	\int De^a_i\!
	     \int DA_i\,\exp\int d^3x\biggl(-\frac{g_3^2}{2}\; e^a_ie^a_i
	+\frac{i}{2}\,\epsilon^{ijk}F^a_{ij}\,e^a_k\biggr).        \la{Z3}
\nnn
\ear
We use the Latin indices $i,j,k\ldots=1,2,3$ to denote spatial directions.
The quantities $e^a_i$ are analogues of the dual field strength
$G^a_{\mu\nu}$ of $d=4$; in three dimensions they are vectors. Similarly to
$d=4$, in addition to invariance under gauge transformations,
\bea
	\delta A^a_i \eql D^{ab}_i\,\alpha^b,
\nn
	\delta e^a_i \eql f^{abc}\,e^b_i\,\alpha^c,
\la{GT3}\eea
the second term in \eq{Z3} is invariant under the
local $(N^2-1)$-function dual gauge transformation
\bea
	\delta A^a_i\eql 0,
\nn
	\delta e^a_i\eql D^{ab}_i(A)\,\beta^b.
\la{DGT3}\eea
As in $d=4$ (see section 2), there is a combination of the transformation
functions $\alpha, \beta$ such that the `bein' $e^a_i$ transforms as a
vector under general coordinate transformations. Therefore, it is
guaranteed that the `mixed' term in \eq{Z3} is diffeomorphism-invariant.
Moreover, along with the 3-function diffeomorphisms, there is an additional
local $(N^2-1-3)$-function symmetry. It has the form of the dual gauge
transformation \ur{DGT3} if one uses the standard form of the BF action but
becomes something extremely interesting when one integrates out the YM
connection $A^a_i$ and writes down the `mixed' term of the action in a
basis-independent form.

The integration over $A^a_i$ is Gaussian. The saddle point $\bar A^a_i$
is found from
\beq
	\epsilon^{ijk}\,D^{ab}_i(\bar A)\,e^b_j=0.
\la{UDgen}\eeq
In the simplest case of the $SU(2)$ gauge group this equation can be
solved in a nice way \cite{Lun1,AMS,DP1}, and we remind it in the next
section.

\section{$SU(2)$, $d=3$ YM theory in gauge-invariant terms}

A general solution of the saddle-point \eq{UDgen} in the case of the
$SU(2)$ gauge group is given by
\beq
	D^{ab}_i(\bar A)\,e^b_j=\Gamma^k_{ij}\,e^a_k,\quad {\rm with}\quad
		\Gamma^k_{ij}=\Gamma^k_{ji}\,.
\la{UD2}\eeq
Following Lunev \cite{Lun1}, we call $e^a_i$ a dreibein and construct
the metric tensor
\beq
	g_{ij}=e^a_ie^a_j\,,\qquad g^{ij}=e^{ai}e^{aj},\quad
			e^a_ie^{bi}=\delta^{ab}\,.
\la{metric3}\eeq
Taking $\d_k\,g_{ij}$ and using \eq{UD2}, one finds that
$\Gamma^k_{ij}$ is the standard Christoffel symbol,
\bearr
    \Gamma_{ij,k}  = \half (\d_i\,g_{jk}+\d_j\,g_{ik}-\d_k\,g_{ij}),\quad
			\Gamma^k_{ij}=g^{kl}\Gamma_{ij,l},
\nnn
\la{Gamma2}\ear
whereas the saddle-point $\bar A^a_i$ is the standard
spin connection made of the dreibein,
\bear
	\bar A^a_i  \eql -\half \epsilon^{abc}\,\omega^{bc}_i,
\nn
	\omega^{bc}_i \eql \half \left[e^{bk}
	(\d_i\,e^c_k-\d_k\,e^c_i) - \,e^{bl}e^{cm}e^a_i \d_le^a_m \right]
\nnn \inch
		-(b\leftrightarrow c).
\la{SC2}\eea
We recall the covariant derivative in curved $d=3$ space,
\beq
	\left(\nabla_i\right)^k_l=\d_i\,\delta^k_l+\Gamma^k_{il},
\la{CD2}\eeq
and build the Riemann tensor which appears to be related to the YM
field strength at the saddle point:
\beq
	\left[\nabla_i\,\nabla_j\right]^k_l=R^k_{\;l\,ij}=
	\epsilon^{abc}\,F^a_{ij}(\bar A)\,e^{bk}\,e^c_l\,.
\la{R2}\eeq
The two terms in the action \ur{Z3} become
\bea
	e^a_i\,e^a_i \eql  g_{ii},
\nn     			\la{twoterms}
	\epsilon^{ijk}\,F^a_{ij}(\bar A)\,e^a_k\eql \sqrt{g}
			R^k_{\;l\,kj}\,g^{jl} =\sqrt{g}\,R.
\eea
Thus, the YM partition function can be rewritten in terms of the local
gauge-invariant variables $g_{ij}$ being the metric of the dual space
\cite{AMS,DP1}:
\beq
	{\cal Z}= \int\! Dg_{ij}\, g^{-\frac{5}{4}}\,\exp\int\!\!
	d^3x\left(-\frac{g^2_3}{2}\,g_{ii}+\frac{i}{2}\,\sqrt{g}\,R\right)\!.
\la{Z23}\eeq
The second term is the Einstein--Hilbert action, while the first term is not
diffeomorphism-invariant, so we call it the `\ae ther' term. The functional
integration measure in \eq{Z23} is obtained as follows \cite{DP1,DP2}:
first, one divides the integration over 9 components of the dreibein $e^a_i$
into integration over three rotations of the dreibein (since the action is
gauge-invariant, one can always normalize the integral over three Euler
angles to unity and cancel it out) and over six components of $g_{ij}$,
\beq
	d^{(9)}e^a_i=d^{(3)}O^{ab}\;d^{(6)}g_{ij}\,g^{-\half}.
\la{intmea1}\eeq
Second, there is another factor of $(\det\, e^a_i)^{-3/2}
=g^{-\frac{3}{4}}$ arising from the Gaussian integration over $A^a_i$.

The `\ae ther' term distinguishes the YM theory from the topological
non-propagating $3d$ Einstein gravity \cite{Wit}; in particular, it is
responsible for the propagation of transverse gluons at short distances
\cite{DP1,DP2}.

The partition function \ur{Z23} is the desired formulation of the
$SU(2)$ YM theory in terms of gauge-invariant variables. We now
generalize it to higher gauge groups.

\section{$SU(N)$, $d=3$ BF gravity in a basis-independent formulation}

As in the $SU(2)$ case, we wish to integrate out the YM connection $A^a_i$
whose saddle-point value is determined by \eq{UDgen} and express the result
in terms of gauge-invariant combinations of $e^a_i$. In total, the
quantities $e^a_i$ carry $3\cdot(N^2-1)$ degrees of freedom, of which
$2\cdot(N^2-1)$ are gauge-invariant or, in other words, basis-independent.
In the $SU(2)$ case these $2\cdot 3=6$ variables are the components of the
metric of the dual space; they can be decomposed into spin 0 (1 dof) and
spin 2 (5 dof's) fields, $1+5=6$.

For $SU(3)$ one needs $2\cdot 8=16$ dof's; these will be the 6
components of the metric tensor $g_{ij}$, plus 10 components of spin 1
(3 dof's) and spin 3 (7 dof's) fields, put together into a symmetric
tensor $h_{ijk}$.

For $SU(4)$ one needs $2\cdot 15=30$ dof's; these will be the previous
16, plus 14 new ones in the form of spin 2 (5 dof's) and spin 4
(9 dof's) fields.

The general pattern is that for the $SU(N)$ gauge group one adds new spin
$N$ and spin $N-2$ fields to the `previous' fields of the $SU(N-1)$ group.
All in all, one has for the $SU(N)$ two copies of spin $2,3\ldots N-2$ and
one copy of the `edge' spins 0, 1, $N-1$ and $N$; they sum up into the
needed $2\cdot(N^2-1)$ dof's.

The invariance of the BF term under the $(N^2-1)$-function dual gauge
transformation \ur{DGT3} translates into an $(N^2-1)$-function local
symmetry which mixes fields with different spins. Below we sketch the
derivation of the appropriate action.

\subsection{The $(N^2-1)$-bein}

It will be convenient to introduce the dual field strength as a
matrix, $e_i=e^a_it^a$, where $t^a$ are $N^2-1$ generators
of $SU(N)$, $\Tr(t^at^b)=\frac{1}{2}\delta^{ab}$, and to rewrite the
saddle-point \eq{UDgen} in the matrix form,
\beq
\epsilon^{ijk}\,[{\cal D}_i(\bar A)\,e_j]=0,\qquad
{\cal D}_i = \d_i -i \bar A^a_i t^a.
\la{UDmatr}\eeq

As noticed in Ref.\,\cite{HJ}, \eq{UD2} is not a general solution of
the saddle-point equation \ur{UDmatr} because the basis $e_i$ is not
complete at $N>2$. Therefore, first of all we have to choose the basis
vielbein, call it $e^a_I$, where both indices run from 1 to $N^2-1$. We
shall use the traceless Hermitian matrices $e_I=e^a_It^a$.

In the $SU(2)$ case we take $e_I=e_i$, with $I=i=1,2,3$ and define the
metric tensor
\beq
	g_{ij}=\Tr\{e_ie_j\}, \qquad g_{ij}g^{jk}=\delta^k_i.
\la{g1}\eeq

For $SU(3)$ we build a quadratic expression in $e_i$ which is a
traceless (both in matrix and gravity senses) rank-2 tensor
\beq
  \ee{i_1i_2}=\frac{1}{2!}\left(\{e_{i_1}e_{i_2}\}-\frac{1}{3}g_{i_1i_2}
			\{e_{k_1}e_{k_2}\}g^{k_1k_2}\right). \la{e2}
\eeq

The symbol $\{\ldots\}$ denotes a sum of all permutations of matrices inside
the curly brackets. There are only 5 independent components of $\ee{i_1i_2}$
since $g^{i_1i_2}\ee{i_1i_2}=0$.  At $N=2$ $\ee{i_1i_2}$ is zero. Thus, the
next five components of $e_I$ are $e_I=\ee{i_1i_2}$ with $I=\{i_1i_2\}
=4,5,6,7,8$. We introduce the gauge-invariant symmetric rank-3 tensor
\beq
	h_{ijk}=\frac{1}{3}\Tr\,\{e_ie_je_k\}=\Tr\,e_i\{e_je_k\};
\la{hijk}\eeq
it describes spin 1 and spin 3 fields and possesses 10 dof's. The spin 1
component is cut out by the contraction, $h_i=h_{ijk}g^{jk}$.

For $SU(4)$ we need further components of $e_I$, namely the irreducible
rank-3 tensor cubic in $e_i$, call it $\ee{i_1i_2i_3}$:
\bearr
	\ee{i_1i_2i_3}=\frac{1}{3!}\left(\{e_{i_1}e_{i_2}e_{i_3}\}
	-\frac{1}{5}g_{i_1i_2}\{e_{k_1}e_{k_2}e_{i_3}\}g^{k_1k_2}\right.
\nnn \quad
	-\frac{1}{5}g_{i_1i_3}\{e_{k_1}e_{i_2}e_{k_3}\}g^{k_1k_3}
	-\frac{1}{5}g_{i_2i_3}\{e_{i_1}e_{k_2}e_{k_3}\}g^{k_2k_3}
\nnn \nhq
\left. -{\bf 1}(h_{i_1i_2i_3} -\frac{1}{5}g_{i_1i_2}h_{i_3}
       -\frac{1}{5}g_{i_1i_3}h_{i_2}-\frac{1}{5}g_{i_2i_3}h_{i_3})\right)\!.
\la{e3}\ear
It has the following properties: a) $\ee{i_1i_2i_3}$ is a Hermitian and
traceless matrix, symmetric under permutations of any indices, b) it has
only 7 independent components since $g^{i_1i_2}\ee{i_1i_2i_3}=0$, c) at
$N=3$ it is automatically zero owing to an identity valid for any
Hermitian $3\times 3$ matrix,
\bear
	M^3\eql M^2\Tr(M)+M\frac{1}{2}\left[\Tr(M^2)-(\Tr\,M)^2\right]
\nnn
	+{\bf 1}\,\det(M).
\la{char32} \eea
Therefore, for $N\geq 4$ we take $e_I=\ee{i_1i_2i_3}$ with
$I=\{i_1i_2i_3\}=9,10, \ldots, 15$.

Using the above recipe for constructing irreducible tensors one can
iteratively build higher-rank tensors suitable for higher groups
and thus higher components of the $(N^2-1)$-bein $e_I$:
\beq
		e_I=(e_i,\,\ee{i_1i_2},\,\ee{i_1,i_2,i_3},\ldots).
\la{Ibein}\eeq
For the general $SU(N)$ group the total number of independent
components of $e_I$ is, as it should be, $\sum_{J=1}^{N-1}(2J+1)=N^2-1$.
\eq{Ibein} reminds the Burnside basis for $SU(N)$.

\subsection{$SU(N)$ metric tensor}

Having built the $(N^2-1)$-bein, we introduce a generalized $SU(N)$ metric
tensor which is a real and symmetric $(N^2-1)\times (N^2-1)$ matrix:
\beq
		g_{IJ}=\Tr\{e_I\,e_J\}.
\la{genmetr}\eeq
It is, generally, not degenerate, therefore one can also introduce the
`contravariant' metric tensor $g^{JK}$, such that
$g_{IJ}g^{JK}=\delta^K_I$. It should be noted that the $\delta$-symbol
here is not the usual Kronecker delta. Although its left-upper-corner
component is the usual $\delta^k_i$, another diagonal component is
\beq
	\delta^{\{k_1k_2\}}_{\{i_1i_2\}}
	=\frac{1}{2}\left(\delta^{k_1}_{i_1}\delta^{k_2}_{i_2}
	+\delta^{k_1}_{i_2}\delta^{k_2}_{i_1}-
	\frac{2}{3}g^{k_1k_2}g_{i_1i_2}\right).
\la{Kr}\eeq

Let us consider the covariant $SU(N)$ metric tensor $g_{IJ}$
in more detail. $I,J$ are multi-indices running $I=(i,\,\{i_1i_2\},\,
\{i_1i_2i_3\},\ldots),\;J=(j,\,\{j_1j_2\},\, \{j_1j_2j_3\},\ldots)$. We
denote
\bearr
	p_{j_1j_2j_3j_4}=\fract{1}{12}\Tr\{e_{j_1}e_{j_2}e_{j_3}e_{j_4}\}
	=\fract{1}{3}\Tr\,e_{j_1}\{e_{j_2}e_{j_3}e_{j_4}\},
\nnnv
	q_{i_1i_2j_1j_2} = \half \Tr\{e_{i_1}e_{i_2}\}\{e_{j_1}e_{j_2}\}.
\la{pq}\ear

We list the first few components of $g_{IJ}$:
\bear
	g_{IJ}\eql g_{ij},\qquad{\rm for}\ \ I=i,\;J=j,
\nn
	g_{i\{j_1j_2\}}  \eql h_{ij_1j_2}-\fract{1}{3}h_ig_{j_1j_2},
\nn
	g_{i\{j_1j_2j_3\}} \eql p_{ij_1j_2j_3}
	-\fract{1}{5}g_{j_1j_2}\,p_{ik_1k_2j_3}\,g^{k_1k_2}
\nnnv \nqq
	- \fract{1}{5}g_{j_1j_3}\,p_{ik_1jk_3}\,g^{k_1k_3}
        - \fract{1}{5}g_{j_2j_3}\,p_{ij_1k_2k_3}\,g^{k_2k_3},
\nn
	g_{\{i_1i_2\}\{j_1j_2\}}\eql q_{i_1i_2j_1j_2}
	-\fract{1}{3} g_{i_1i_2}\,q_{k_1k_2j_1j_2}\,g^{k_1k_2}
\nnn \nq
	-\fract{1}{3}g_{j_1j_2}\,q_{i_1i_2k_1k_2}\,g^{k_1k_2}
\nnn \nq
	+\fract{1}{9}g_{i_1i_2}\,g_{j_1j_2}q_{k_1k_2l_1l_2}
		\,g^{k_1k_2}\,g^{l_1l_2},
\la{compg}\ear
and so on.

An important question is that of the number of independent degrees of freedom
encoded in various components of $g_{IJ}$. The symmetric tensor $g_{ij}$
contains 6 dof's; it can be decomposed into spin 2 and spin 0 fields. For
SU(2), six is exactly the number of gauge-invariant dof's. Next,
$g_{i\{j_1j_2\}}$ has 10 dof's which can be viewed as those belonging to
spin 3 and spin 1 fields. The latter is represented by the vector field
$h_i$, the former is represented by the symmetric and traceless rank-3
tensor written in the last line of \eq{e3}. The $(6+10)=16$ dof's of
$g_{ij}$ and $g_{i\{j_1j_2\}}$ together compose the needed $2(3^2-1)=16$
gauge-invariant dof's of the $SU(3)$ group. It means, in particular,
that the components $g_{\{i_1i_2\}\{j_1j_2\}}$ are not independent
variables but are algebraically expressible through the tensors
$g_{ij}$ and $h_{ijk}$.

Let us consider the $SU(4)$ case. One has to add the components
$g_{i\{j_1j_2j_3\}}$ which are, generally speaking, a mixture of spins
4,2 and 0. However, spin 0 is in fact absent in this tensor. To see it,
we contract it with the combination
\beq
	\fract{1}{3}\left(g^{ij_1}g^{j_2j_3}+g^{ij_2}g^{j_1j_3}
		+g^{ij_3}g^{j_1j_2} \right) \la{zerospin}
\eeq
which cuts out the spin 0 component of the
rank-4 tensor. This contraction is zero, demonstrating that
spin 0 is absent. Therefore, $g_{i\{j_1j_2j_3\}}$ can be decomposed into
spin 4 and spin 2 components only and thus carries
$(2\cdot 4+1)+(2\cdot 2 +1)=9+5=14$ dof's. These 14 add up with the
previous 16 to give exactly 30 dof's coinciding with the number of
gauge-invariant combinations for the $SU(4)$ group. Acting in the same
fashion, one can verify that the component $g_{i\{j_1j_2j_3j_4\}}$
arising for groups $SU(5$) (and higher) has only spins 5 and 3 but not
spin 1 and thus contains $11+7=18$ dof's which, together with the
previous 30 give the 48 gauge-invariant variables of the $SU(5)$ group.

It means that the first line in the general metric tensor, namely
$g_{iJ}$ with $i=1,2,3,\;J=1\ldots N^2-1$, contains exactly $2\cdot(N^2-1)$
dof's, i.e., the needed amount; all the rest components are algebraically
expressible in terms of the first line (or the first column, $g_{Ij}$).

\subsection{$SU(N)$ generalization of the Einstein-Hilbert action}

We are now equipped for seeking a solution of the saddle-point
\eq{UDmatr} in the form generalizing \eq{UD2} to an arbitrary gauge group
\bear
	[{\cal D}_i\,e_j]\eql \Gamma^k_{ij}e_k+\Gamma^{\{k_1k_2\}}_{ij}
	\ee{k_1k_2}
\nnn \nq
	+\Gamma^{\{k_1k_2k_3\}}_{ij}\ee{k_1k_2k_3}+\ldots\eqdef
			\Gamma^K_{ij}\,e_K.
\la{UDG1}
\eea
It solves \eq{UDmatr} provided the Christoffel symbols are symmetric,
$\Gamma^K_{ij}=\Gamma^K_{ji}$. It is important that $N^2-1$ components of
$e_K$ form a complete set of Hermitian and traceless $N\times N$ matrices,
therefore \eq{UDG1} gives a general solution of \eq{UDmatr}.

By considering derivatives of the metric tensor $g_{iI}$ and using
\eq{UDG1}, one can express all $\Gamma^K_{ij}$ in terms of the
gauge-invariant (i.e., basis-independent) variables $g_{iJ}$ and their
derivatives. Moreover, since all higher components of $e_I$ are
explicitly constructed from the first three $e_i$'s it is possible
to generalize \eq{UDG1} introducing the generalized Christoffel symbols
$\Gamma^K_{iJ}$ such that
\beq
		[{\cal D}_ie_J]=\Gamma^K_{iJ}e_K.
\la{genChr}\eeq
The generalized Christoffel symbols are found from the condition that the
covariant derivative of $g_{IJ}$ is zero,
\beq
     g_{IJ;\,k}=\d_k\,g_{IJ}-\Gamma^L_{kI}\,g_{LJ}-\Gamma^L_{kJ}\, g_{IL}=0,
\la{covderz}\eeq
which expresses the $\Gamma$'s through the derivatives of the metric
tensor. This equation follows from considering the derivative
$\d_k\,\Tr\{e_Ie_J\}=\Tr\{[{\cal D}_ke_I]e_J\}+
\Tr\{e_I[{\cal D}_ke_J]\}$ and using \eq{genChr}.

We next consider the double commutator,
\bear
	[{\cal D}_m[{\cal D}_n e_I]] \eql [{\cal D}_m,\Gamma^K_{nI} e_K]
\nnn \ \
	=\d_m\Gamma^K_{nI}e_K+\Gamma^J_{mK}\Gamma^K_{nI}e_J.
\la{doublecom}\ear
We interchange $(mn)$ and subtract one from another:
\bearr          \nhq
	[{\cal D}_m[{\cal D}_n e_I]]-[{\cal D}_n[{\cal D}_m e_I]]
		=-[e_I[{\cal D}_m{\cal D}_n]]=i[F_{mn}e_I]
\nnn \qquad
	=\left(\d_m \Gamma^J_{nI}-\d_n\Gamma^J_{mI}+
	\Gamma^J_{mK}\Gamma^K_{nI}-\Gamma^J_{nK}\Gamma^K_{mI}\right)e_J
\nnn \qquad
	=[\nabla_m\,\nabla_n]^J_I\,e_J	\eqdef R^J_{I\,mn}\,e_J,
\nnn
	(\nabla_m)^J_K \eqdef  \delta^J_K\,\d_m+\Gamma^J_{mK},
\la{Riemann}\ear
which serves as a definition of the generalized Riemann tensor and
simultaneously expresses the Yang--Mills field strength $F_{mn}$ at the
saddle point in terms of the Riemann tensor. In the component form
\eq{Riemann} is
\beq
	-f^{abc}F^a_{mn}e^b_I=R^J_{I\,mn}\,e^c_J.
\la{fs1}\eeq
We contract this equation with the contravariant vielbein $e^{dI}$,
$e^b_Ie^{dI}=\delta^{bd}$,
\beq
		-f^{abc}F^a_{mn}\delta^{bd}=R^J_{I\,mn}\,e^c_J\,e^{dI},
\la{fs2}\eeq
and then with $f^{cde}$,
\beq
		N\,F^e_{mn}=f^{cde}\,e^c_J\,e^{dI}\,R^J_{I\,mn}.
\la{fs3}\eeq

The action density of the mixed term in \eq{Z3} is
\beq
	{\cal L}= \epsilon^{mnp}F^a_{mn}e^a_p=\frac{1}{N}\epsilon^{mnp}
		f^{cde}\,e^c_J\,e^{dI}\,e^e_p\,R^J_{I\,mn},
\la{lagr}\eeq
so that the BF action becomes an $SU(N)$ generalization of
the Einstein-Hilbert action,
\bea
	S_2\eql \frac{i}{2}\int d^3x\,{\cal L}
\nn
	\eql \frac{1}{N}\int d^3x\,\Tr\left([e_Ie^J]e_p\right)\,
			R^I_{J\,mn}\,\epsilon^{mnp}.
\la{act2}\eea
The first factor here is gauge-invariant (i.e., basis-independent) and is
an algebraic combination of the components of the metric tensor $g_{iJ}$.
The Riemann tensor is also made of $g_{iJ}$ and its derivatives and is
thus gauge-invariant as well.

In the $SU(2)$ case when $I=i=1,2,3$ and $J=j=1,2,3$, \eq{act2} can be
simplified since
\beq
	\Tr\left([e_ie^j]e_p\right)=\frac{i}{2}\epsilon_{ij' p}\,
		g^{jj'}\,\sqrt{g},
\eeq
so that we obtain the standard Einstein--Hilbert action,
\beq
	S_2\Big|_{SU(2)} = \frac{i}{2}\int\! d^3x\, \sqrt{g}\, R.
\eeq

Finally, the first term in \eq{Z3} is simply
\beq
		S_1=-\frac{g^2_3}{2}\int d^3x\, g_{ii},
\la{act1}\eeq
just as in the $SU(2)$ case. We note that the relative strength of the two
terms is $g^2_3 N$, which has a finite limit at $N\to\infty$.

\eqs{act2}{act1} solve, in a somewhat symbolic form, the problem of
rewriting an arbitrary Yang--Mills theory in $d=3$ in terms of
$2\cdot(N^2-1)$ gauge-invariant variables contained in the metric
tensor $g_{iJ}$. However, there remains an algebraic problem of
expressing explicitly all quantities described in this subsection
through $g_{iJ}$ and its derivatives.

\subsection{Local symmetry mixing fields with different spins}

We will find here a whole new class of local transformations mixing
fields with different spins. In contrast to supergravity, only boson
fields are involved, unless one considers the generalization of the BF
action to incorporate supersymmetric fermions \cite{Sch}.

This new symmetry originates, of course, from the symmetry of the mixed term
in the first-order formalism (or of BF gravity) under dual gauge
transformations \ur{DGT3} and reveals itself when one integrates out the YM
connection $A_i$. We decompose the infinitesimal matrix of the dual gauge
transformation $\beta$ in the $(N^2-1)$-bein $e_K$, $\beta=y^Ke_K$, and use
the saddle-point equation (\ref{genChr}) to rewrite the transformation
in terms of covariant derivatives of the $N^2-1$ infinitesimal functions
$y^K$. The variation of the vielbein under which the action is invariant is
\bea
	\delta\,e_i \eql  [{\cal D}_i(\bar A) \,\beta]
		= (\d_i\,y^K)e_K+y^K\Gamma^L_{iK}\,e_L
\nn
	\eql e_L\left(\nabla_i\right)^L_K\,y^K.
\la{DGTnew}\eea
The variation of the $(ij)$ component of the metric tensor
\ur{genmetr} is, consequently,
\bearr
	\delta\,g_{ij}=\delta\,\Tr(e_ie_j)
	=g_{iL}\left(\nabla_j\right)^L_K\,y^K + g_{Lj}(\nabla_i)^L_K\,y^K
\nnn
	=g_{iK}\d_jy^K+g_{Kj}\d_iy^K
		+\left(g_{iL}\Gamma^L_{jK}+g_{Lj}\Gamma^L_{iK}\right)y^K.
\nnn
\la{transf1}\ear
Similar variations can be found for the other components of the
generalized metric tensor $g_{IJ}$.

In the particular case of diffeomorphisms one takes only three functions
$y^K(x)$ with $K=k=1,2,3$ and puts the rest $N^2-4$ functions to be zero.
Using \eq{covderz} one sees that in this case the variation \ur{transf1}
becomes the usual general coordinate transformation of the metric tensor,
\beq
	\delta\,g_{ij}=g_{ik}\,\d_jy^k+g_{kj}\,\d_iy^k + \d_ k g_{ij}\,y^k.
\la{diff1}\eeq
In the $SU(2)$ case this is the only symmetry we have; the BF action
\ur{act2} can be written as $\int \sqrt{g} R$, and its only
local symmetry is the invariance under diffeomorphisms.

For higher groups there is an additional symmetry. In the general case
when all $N^2-1$ functions $y^K(x)$ are nonzero the last term in
\eq {transf1} is a combination of the derivatives of the generalized
metric tensor components $g_{IJ}$ which, are, in turn, algebraic
combinations of the $2\cdot(N^2-1)$ independent variables $g_{iJ}$.

We thus see that the BF action \ur{act2} is invariant under $N^2-1$ local
transformations which mix, in a nonlinear way, various components of the
metric tensor $g_{iJ}$. For example, in case the gauge group is $SU(3)$, the
transformation \ur{transf1} mixes the gauge-invariant fields $g_{ij}$ and
$h_{ijk}$ carrying spins 0, 1, 2 and 3. In case of the $SU(N)$ gauge group
the transformation \ur{transf1} mixes fields with spin from 0 to $N$.  At
$N\to\infty$ the BF action \ur{act2} has an infinite-dimensional local
symmetry and mixes an infinite tower of spins, with all (integer) spins
twice degenerate, except the lowest spins 0 and 1.

It reminds the symmetry of string theory. It would be interesting to
reformulate the material of this section directly in the $N\to\infty$
limit and to reveal this symmetry explicitly. It must be similar but not
identical to the Virasoro algebra whose realization is physical only in 26
dimensions. Neither is it the $W_{1+\infty}$ algebra introduced in
connection with the inclusion of higher spins into general relativity in
Ref.\,\cite{BdWV}, since the spin content of the $SU(N)$ Yang--Mills theory
is different.

We notice that the `\ae ther' term \ur{act1} is the `dilaton' term; in
$d=2+1$ the dimension of the gauge coupling constant $g_3^2$ is that of a
mass. Therefore, one expects that the role of this term is to lift the
degeneracy of the otherwise massless fields and to provide the string with a
finite slope $\alpha'=O\left((g_3^2N)^{-2}\right)$.

\section{Conclusions}

Using the first-order formalism as a starting point, we have reformulated
the $SU(N)$ Yang--Mills theory in $d=3$ and the $SU(2)$ theory in $d=4$ in
terms of local gauge-invariant variables.  In all cases these variables are
either identical or closely related to the metric of the `colour dual
space'. The Yang--Mills action universally contains two terms: one is a
generalization of the Einstein--Hilbert action and possesses large local
symmetry which includes invariance under diffeomorphisms, the other (`\ae
ther') term does not have this symmetry but is simple. The `\ae ther' term
distinguishes Yang--Mills theory from non-propagating topological BF
gravity.

BF gravity based on a gauge group $SU(N)$ is known to possess invariance
under dual gauge transformations characterized by $N^2-1$ functions in
$d=3$ and by $4\cdot (N^2-1)$ functions in $d=4$. This symmetry is
apparent when BF theory is presented in terms of the YM connection
but is not so apparent when one integrates over the connection (which
is possible since the integral is Gaussian) and writes down the result
of the integration in basis-independent variables, e.g., the metric tensor.
We have shown that in such a case the original invariance under dual gauge
transformations does not disappear but manifests itself as a symmetry under
(generally, a nonlinear one) mixing fields with different spins. The higher
is the gauge group, the higher spins transform through one another.  At
$N\to\infty$, an infinite tower of spins are related by symmetry
transformation. Since it is the kind of symmetry known in string theory, and
some kind of string is expected to be equivalent to the Yang--Mills theory
in the large $N$ limit, it is tempting to use this formalism as a starting
point for deriving a string from a local field theory.

\Acknow
{We are grateful to Ian Kogan for stimulating remarks. Part of this
work has been done during the authors' participation in an INT
programme in Seattle, and we thank INT for hospitality. We acknowledge
partial support by the RFBR grant 00-15-96610.}

\end{document}